# Intrinsic and extrinsic conduction contributions at nominally neutral domain walls in hexagonal manganites

*J. Schultheiß[1], J. Schaab[2], D.R. Småbråten[1], S.H. Skjærvø[1], E. Bourret[3], Z. Yan[2,3], S.M. Selbach[1], and D. Meier[1,*]*

[1] Department of Materials Science and Engineering, NTNU Norwegian University of Science and Technology, 7034 Trondheim, Norway
[2] Department of Materials, ETH Zurich, 8093 Zurich, Switzerland
[3] Materials Science Division, Lawrence Berkeley National Laboratory, Berkeley, CA, USA

*corresponding author: dennis.meier@ntnu.no

**Conductive and electrostatic atomic force microscopy (cAFM and EFM) are used to investigate the electric conduction at nominally neutral domain walls in hexagonal manganites. The EFM measurements reveal a propensity of mobile charge carriers to accumulate at the nominally neutral domain walls in ErMnO$_3$, which is corroborated by cAFM scans showing locally enhanced d.c. conductance. Our findings are explained based on established segregation enthalpy profiles for oxygen vacancies and interstitials, providing a microscopic model for previous, seemingly disconnected observations ranging from insulating to conducting domain wall behavior. In addition, we observe variations in conductance between different nominally neutral walls that we attribute to deviations from the ideal charge-neutral structure within the bulk, leading to a superposition of extrinsic and intrinsic contributions. Our study clarifies the complex transport properties at nominally neutral domain walls in hexagonal manganites and establishes new possibilities for tuning their electronic response based on oxidation conditions, opening the door for domain-wall based sensor technology.**

Ferroelectric domain walls are natural interfaces that offer exciting opportunities for fundamental research and a rich potential for applications. The dynamic characteristics of ferroelectric domain walls introduce spatial mobility, allowing the real-time manipulation of their position, configuration and density.[1,2] This mobility adds a degree of flexibility which is not available at conventional interfaces, enabling the domain walls to take an active role in next-generation devices. In addition, individual domain walls can be used to emulate the behaviour of basic electronic components, such as diodes,[3] memories[4] or transistors,[5] and thus hold great promise as functional 2D systems.[6,7]

Despite the tremendous progress in experiment and theory, the microscopic details behind many of the emergent electronic domain wall properties remain to be understood. It is now clear that both intrinsic effects (e.g., band bending[8] and polar discontinuities[9]) and extrinsic effects (defects[10,11]) can give rise to unusual direct current (d.c.) conductance at ferroelectric domain walls. While at charged domain walls, the diverging electrostatic potentials enhance conductance (intrinsic),[12,13] nominally neutral domain walls do not display an electric field driven accumulation of charge carriers. Other effects, however, such as local elastic strain fields, can affect the conduction properties at neutral domain wall. For instance, oxygen vacancies have been found to accumulate and give rise to increased conductance at nominally neutral domain walls.[14-16] For most ferroelectric domain walls, however, point defect densities are usually unknown and different conduction contributions intermix, which complicates the situation. The latter is strikingly reflected by the complex transport behaviour at nominally neutral ferroelectric domain walls in hexagonal manganites, for which electronic properties were found to range from insulating to highly conducting.[13,17-27] Theoretical approaches agree that the transport at the nominally neutral domain walls is due to extrinsic effects and governed by oxygen defects, but a comprehensive model is virtually non-existent. Atomistic simulations in YMnO$_3$, for example, find that oxygen vacancies accumulate at neutral domain walls[28], whereas density functional theory (DFT) calculations predict that oxygen vacancies are more stable in the bulk.[29] Recently, it was further calculated that oxygen interstitials have a propensity towards accumulation at neutral domain walls.[24]

Here, to clarify the domain wall conductivity at neutral domain walls in hexagonal manganites and develop a comprehensive model, we investigate ErMnO$_3$. ErMnO$_3$ displays improper ferroelectricity with a spontaneous polarization $P$ of ≈6 μC/cm² at room temperature (space group symmetry $P6_3cm$), with $P$ parallel to the crystallographic c-axis.[30,31] In our work, we identify nominally neutral domain walls on c-oriented ErMnO$_3$ single-crystals by piezoresponse force microscopy (PFM). The presence of mobile charge carriers at these domain walls is demonstrated by electrostatic force microscopy (EFM). Complementary conductive atomic force microscopy (cAFM) measurements confirm enhanced domain wall conductance. Furthermore, we observe significant variations in the conductance at nominally neutral domain walls, which we attribute to sub-surface deviations from the ideal neutral structure and associated domain wall bound charges. Our findings corroborate the electronic changeability of neutral domain walls in hexagonal manganites, ranging from insulating to conducting states, which we explain within the framework of established segregation enthalpies.[24,29] This changeability is a direct consequence of the outstanding chemical flexibility of this material class, readily hosting both oxygen interstitials and vacancies that interact with the neutral domain walls.

For our experimental studies, high-quality ErMnO$_3$ single crystals are grown by the pressurized floating-zone method.[32] Laue diffraction is used to orient the crystal with $P$ pointing out-of-plane to achieve a surface with nominally neutral domain walls. The samples with a thickness of about 1 mm are then cut and lapped with a 9 μm-grained Al$_2$O$_3$ water suspension and polished using silica slurry (Ultra-Sol® 2EX, Eminess Technologies, Scottsdale, AZ, USA) to produce a flat surface with a root-mean-square roughness of about 6.3 nm (averaged over a 10x10μm² scan area). The out-of-plane orientation of the ferroelectric polarization in our ErMnO$_3$ sample is confirmed by calibrated PFM. Maps of the mobile charge carrier density and conductance are obtained by EFM and cAFM, respectively. EFM microscopy measurements are performed on a Ntegra Prima system in combination with a SMENA head (NT-MDT, Moscow, Russia), while for PFM and cAFM measurements a SF005 head was used. A Pt-Ir conductive ANSCM-PT tip (ANSCM-PT, AppNano, Maintain View, CA, USA) is used in contact mode with SF005 head (NT-MDT, Moscow, Russia) for PFM and cAFM measurements. PFM measurements were carried out at 5.0 V and 41.35 kHz. For EFM measurements DCP20 tips (NT-MDT, Moscow, Russia) are used in non-contact mode with 6.0 V applied at 37.09 kHz.

We begin our analysis by probing the electronic response at nominally neutral domain walls in non-contact mode using EFM. A representative scan is displayed in Figure 1a. The 2ω channel is read out, which quantifies the polarizability / mobile charge carriers.[33] A calibrated PFM phase image taken at the marked position is presented, displaying the spatial variation of the out-of-plane polarization vector. The PFM data identifies the positions of the domain walls, which correlate with the enhanced contrast seen in EFM. The EFM and PFM scans suggest an increased density of mobile charge carriers at the nominally neutral domain walls relative to the surrounding domains. To further elaborate on the enhanced density of mobile carriers detected by EFM, we perform local conductance measurements using cAFM in contact mode (Figure 1b). Enhanced contrast in the cAFM data indicates higher conductance and can be clearly correlated to the position of the nominally neutral domain walls. We note that – in contrast to previous reports[24,26] - the transport measurement at the domain walls is free from additional domain-related signals. Domain-related cAFM signals have previously been observed in hexagonal manganites with out-of-plane polarization orientation due to domain-modulated Schottky-like rectification of injected currents.[26] A statistical analysis of the cAFM data in Figure 1b is presented in Figure 1c, showing that the conductance at the domain walls ($I_{DW}$=0.15 pA) is significantly enhanced in comparison to the domains ($I_D$=0.01 pA). In agreement with the spatially resolved cAFM data, voltage-dependent current measurements (supplementary Figure S1a) reflect a significant variation in the conductance behavior at the nominally neutral domain walls. Time-dependent conductance measurements (supplementary Figure S1b) exclude scan artefacts and spurious contributions due to the switch-on-capacitance, rearrangement of bound charge carriers or the movement of domain walls as possible origin.[10,34,35]

Figure 1 demonstrates enhanced d.c. conductance at the nominally neutral domain walls. Furthermore, on a closer inspection, the EFM and cAFM data in Figure 1 reveal variations in the transport behavior of different domain walls. This effect is highlighted in Figure 2a and b, where both EFM/cAFM and PFM signals are plotted along the dashed line in Figure 1a and b, respectively. The enhanced contrast in the EFM/cAFM measurements correlates with the change in PFM phase contrast, with alternating high and low amplitude of the EFM/cAFM signal. This variation is surprising as all domain walls are nominally neutral on this c-oriented ErMnO$_3$ sample with out-of-plane polarization. It is important to note, however, that while EFM and cAFM are sensitive to surface effects, they also probe surface-near regions. Accounting for the fact that we are recording domain wall contributions from within the bulk, too, the variations in conductance can be understood based on the realistic assumption that the walls are not perfectly parallel to $P$ as schematically illustrated in Figure 2c. In general, as already discussed in ref. 24, extrinsic contributions, e.g., from point defects, can lead to enhanced conductance. This enhancement is expected to be equally pronounced for all neutral domain wall, denoted as $I_0$ in Figure 2c. However, at positions where the walls tilt away from the ideal charge-neutral domain wall structure and form highly conducting tail-to-tail walls,[13] additional intrinsic conduction contributions, $\Delta I$, arise as schematically depicted in Figure 2c. Values obtained from cAFM measurements (Figure 2b, $I_0$=$\Delta I$=0.1pA) suggest that extrinsic (point defects) and intrinsic (hole accumulation due to sub-surface tail-to-tail configuration) current contributions at nominally neutral domain walls can be equally pronounced. This result shows the importance of "hidden" sub-surface effects in domain wall studies when aiming for quantitative insight. Possible experimental strategies may involve FIB (Focused Ion Beam) techniques, creating ultra-thin samples with well-defined domain wall states[36,37] or bulk-sensitive diffraction.[38,39]

Our experimental data corroborate that nominally neutral domain walls in hexagonal manganites can exhibit enhanced d.c. conductance, consistent with previous transport measurements.[21-25] This d.c. conductance is commonly attributed to extrinsic effects that redistribute due to local stain fields, but wall-to-wall variations may arise due to additional intrinsic contributions. The latter are associated with the meandering nature of domain walls within the bulk[24,40], which leads to deviations from the ideal neutral domain wall structure (Figure 2c).

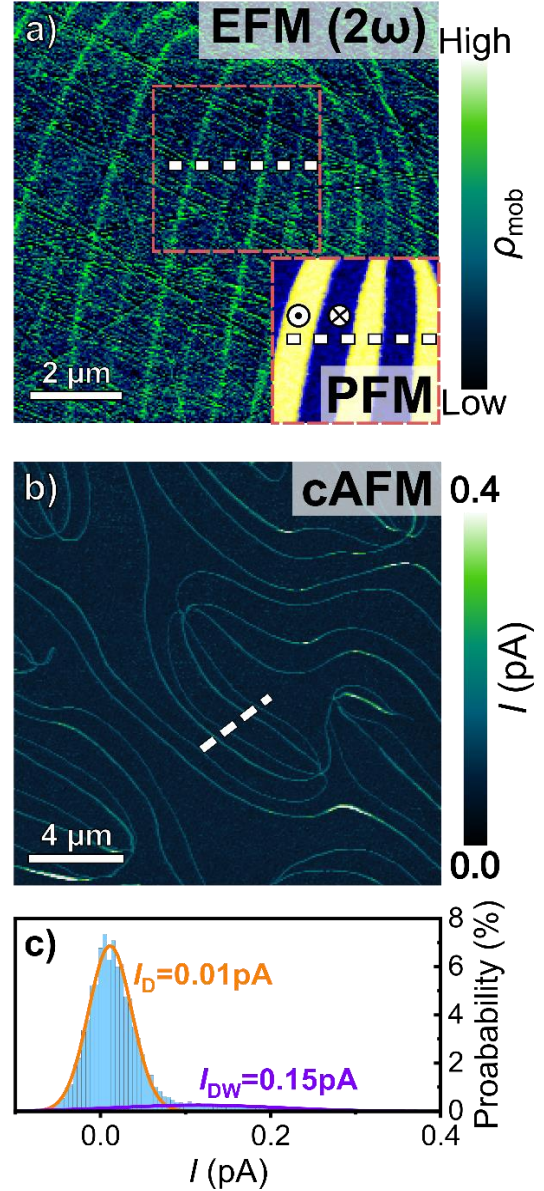

**Figure 1:** a) EFM contrast (2ω signal) recorded on an ErMnO$_3$ single crystal with out-of-plane polarization, revealing the density of mobile charge carriers, $\rho_{mob}$. PFM phase contrast for the red square is displayed in the inset. Up- and downward polarization is identified by yellow and blue areas, respectively. The EFM signal in a) indicates increased response at the nominally neutral domain walls, which can be attributed to an enhanced concentration of mobile charge carriers. b) cAFM map gained on ErMnO$_3$ with out-of-plane polarization. The cAFM image (taken at 2.0V) displays enhanced d.c. conductivity at the nominally neutral domain walls. c) Histogram displaying the probability of current values obtained from the cAFM scan in b). The solid lines represent Gaussian fits of the probability distributions with average currents for domain wall ($I_{DW}$) and domain current ($I_D$), as indicated.

After discussing the enhanced d.c. conductance observed at nominally neutral domain walls in our ErMnO$_3$ crystal, we now turn to the wide range of conduction properties for neutral domain walls in hexagonal manganites reported in literature.[13,17-27] We note that while strain fields at neutral domain walls do not significantly impact their intrinsic conductivity,[41] they are critical for the local defect chemistry. In general, whenever the local strain field around a specific point defect is similar to that at domain walls, local minimization of the total energy will promote segregation of this type of defect, leading to an accumulation at the domain walls.[24, 29,42,43] In contrast, different local strain fields are expected to promote a depletion of point defects at domain walls. Taking such strain effects into account, Figure 3 summarizes previously reported oxygen point defect segregation enthalpies. Segregation enthalpy profiles for oxygen vacancies[29] ($V_O$) and oxygen interstitials[24] ($O_i$) in the structurally distorted volumes in the vicinity of the walls[40] are displayed in Figure 3a and b, respectively. On the one hand, the results reveal that $V_O$ segregate away from the neutral walls, with a positive segregation enthalpy of +0.12 eV relative to the bulk.[29] Note that the oxygen vacancy defect chemistry of $YMnO_3$ is dominated by $V_{O4}$,[29,42,43] which has a 10-20% lower formation energy compared to the other oxygen vacancies.[29] On the other hand, oxygen interstitials tend to segregate towards neutral walls with a negative segregation enthalpy of -0.03 eV relative to the bulk.[24]

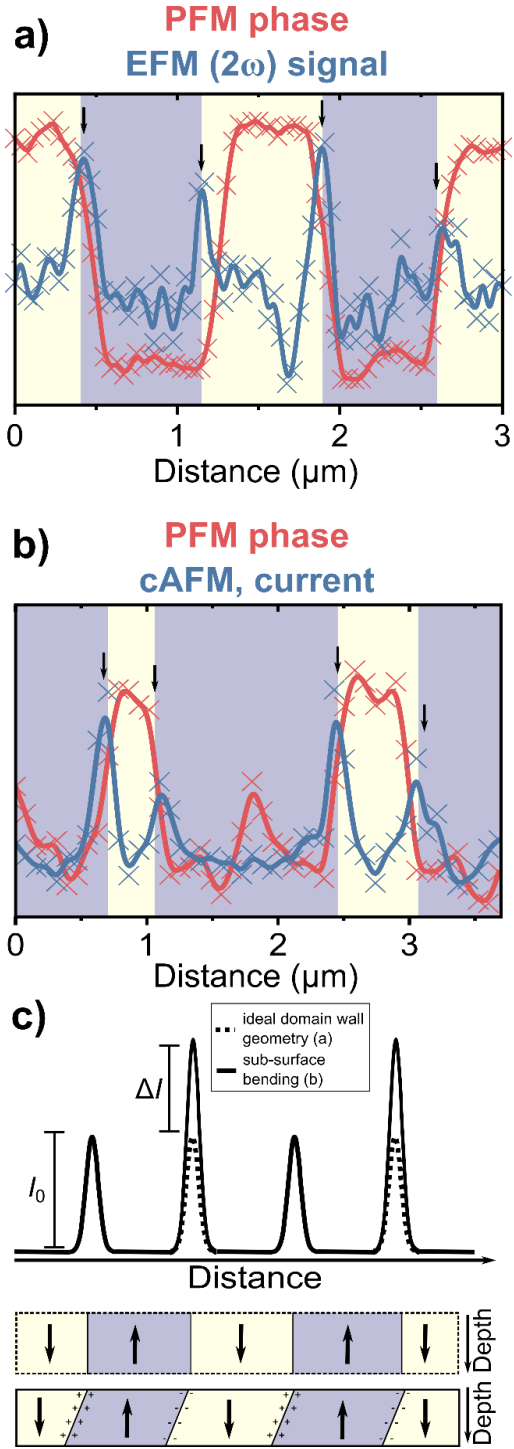

**Figure 2:** a) EFM and b) cAFM signals are displayed simultaneously with PFM phase contrast, corresponding to the dashed lines in Figure 1a and 1b, respectively. Up- and downward polarization is identified by yellow and blue colors, respectively. The small black arrows indicate the position of nominally neutral domain walls, highlighting the observation of alternating high and low currents between the walls. c). Schematic illustration of the expected (dashed line) and measured (solid line) current contribution for a sequence of nominally neutral domain walls. Due to the accumulation of oxygen interstitials,[24] neutral walls are expected to exhibit enhanced extrinsic conductance, $I_0$, relative to the surrounding bulk. Deviations from the ideal charge-neutral structure within the bulk, however, give rise to domain wall bound charges as indicated by + and − symbols. In case of negative bound charges (−), corresponding to the formation of a tail-to-tail domain wall within the bulk, it is established that mobile holes accumulate, which further increases the local intrinsic conductance by ∆$I$. Values of $I_0$=∆$I$=0.1pA were identified from Figure 2b.

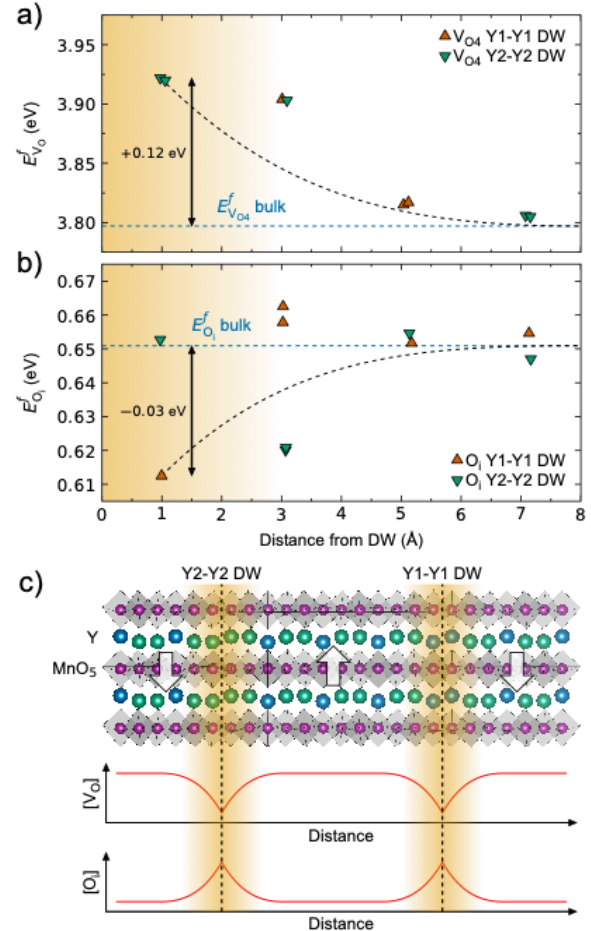

**Figure 3:** Segregation enthalpies for a) the most stable type of oxygen vacancies[29] ($V_O$) and b) oxygen interstitials[24] ($O_i$) are displayed as a function of the distance from the neutral domain wall (DW). The yellow area indicates the extension of the structural distortion from the domain wall into the bulk.[40] The segregation enthalpy for $V_O$ is positive, while the segregation enthalpy for $O_i$ is negative towards the neutral domain wall, which indicates a depletion of $V_O$ and an accumulation of $O_i$ relative to the surrounding domain. c) Illustration of an $YMnO_3$ supercell used for DFT calculations. White arrows illustrate the polarization direction for each domain. Blue and green Y-atoms correspond to the Y1 and Y2 sites in the $P6_3cm$ space group, respectively.[45] The vertical dashed lines indicate the positions of the two types of neutral domain walls arising from periodic boundary conditions, Y1-Y1 and Y2-Y2, as differentiated by their Y termination at the walls.[41,46] Note that in a real system, neutral domain walls show a continuum between Y1-Y1 and Y2-Y2 termination.[41] The corresponding concentration profiles of [$V_O$] and [$O_i$] are schematically sketched across the super cell. Near the domain wall, [$O_i$] is enhanced, while [$V_O$] is reduced. Note that $V_O$ and $O_i$ are positioned in-plane with Mn.[24,29]

Resulting schematic concentration profiles of oxygen vacancies ([$V_O$]) and oxygen interstitials ([$O_i$]) along the $YMnO_3$ supercell are shown in Figure 3c. The profiles indicate a reduced concentration of vacancies and an enhanced concentration of interstitials at the two neutral walls. Thus, for oxygen hyper-stoichiometry ($YMnO_{3+\delta}$ with $\delta > 0$, through oxygen interstitials[44]) oxygen interstitials accumulate at the neutral walls. Local charge compensation occurs via two holes,[44] which are responsible for the enhanced p-type conductivity at the neutral domain wall (Figure 1). In contrast, in the case of oxygen deficiency ($\delta < 0$, through oxygen vacancies[29]), oxygen vacancies accumulate inside the domains. The latter leads to increased bulk conductivity (n-type), whereas the neutral domain walls remain insulating. With this, the calculated oxygen segregation enthalpies summarized in Figure 3, provide a coherent explanation for the wide range of transport behavior observed at neutral domain walls in hexagonal manganites.[13,17-27]

In summary, our study establishes that mobile charge carriers can accumulate at nominally neutral domain walls in hexagonal manganites, giving rise to enhanced d.c. conductance. EFM and cAFM measurements on high-quality ErMnO$_3$ single crystals reveal significant variations in the conductance at nominally neutral domain walls, which we attribute to a tilting away from the ideal charge-neutral wall structure in surface near regions. Based on a systematic analysis of segregation enthalpy profiles for oxygen vacancies and interstitials from the literature, we develop a comprehensive microscopic model for the local transport behavior. Our microscopic model provides a coherent explanation for the diverse conduction properties observed at nominally neutral domain walls – ranging from insulating to conducting – and advances the fundamental understanding of their complex nanoscale physics. The observed interplay of extrinsic and intrinsic conduction contributions offers new opportunities for property engineering at ferroelectric domain walls, exploiting both their chemical and orientational degrees of freedom. In particular, the correlation between oxygen stoichiometry and conductivity enables the development of ultra-small environmental sensors, where variations in oxygen partial pressure are detected by monitoring the domain wall conductance.

See the supplementary material for voltage and time-dependent current cAFM measurements at neutral domain walls and inside the domains (supplementary material Fig. S1).

J.Schultheiß acknowledges the support of the Alexander von Humboldt Foundation through the Feodor-Lynen fellowship. D.M. and J.Schaab acknowledge funding from the SNF (Proposal No. 200021_149192). D.M. thanks NTNU supported through the Onsager Fellowship Program and the Outstanding Academic Fellows Program. D.R.S., S.H.S., and S.M.S. acknowledge The Research Council of Norway (FRINATEK project no. 231430/F20 and 275139/F20) and NTNU for financial support. Computational resources were provided by UNINETT Sigma2 (project no. NN9264K and ntnu243) and the Euler cluster at ETH Zürich. Z.Y. and E.B. were supported by the U.S. Department of Energy, Office of Science, Basic Energy Sciences, Materials Sciences and Engineering Division under Contract No. DE-AC02-05-CH11231 within the Quantum Materials program-KC2202.

**Supplementary material**

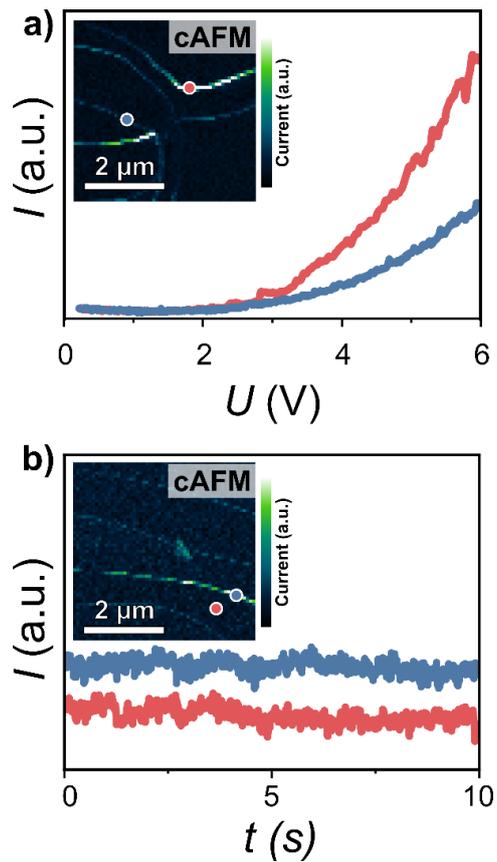

**Figure S1:** a) I(V) curves at the indicated nominally neutral DWs with a high (red dot) and low (blue dot) conductivity. b) Time-dependence of the current recorded at a neutral domain wall (blue dot) and inside (red dot) the adjacent domain. The curves are displayed after the stabilization from currents resulting from switching on the voltage. The inset displays cAFM images (recorded at 2.5V) to identify the measurement positions.